\documentstyle{article}

\begin{document}
\rightline{Nuovo Cimento B 113 (Nov. 1998) 1431-1436}
\rightline{gr-qc/9804047}
\begin{center} {\Large {\bf USING STRICTLY ISOSPECTRAL
UNBROKEN NONRELATIVISTIC SUPERSYMMETRY AS A TOY MODEL}\\
\vskip 3mm
H. Rosu
\footnote{FAX: +0052-47-187611; e-mail: rosu@ifug3.ugto.mx}
}

{\scriptsize Instituto de F\'{\i}sica - IFUG,
Apdo Postal E-143,
Le\'on, Gto, Mexico}
\end{center}
{\bf Summary}: I employ heuristically the strictly isospectral double Darboux
method based on the general superpotential of unbroken nonrelativistic
supersymmetry suggesting a few small steps of principle for extending its
range of applications toward relativistic (gauge) physics. The application of
the method to minisuperspace quantum cosmology is also briefly presented.


\bigskip
\noindent
PACS: 11.30.Pb - Supersymmetry.\\[0.2cm]
PACS: 04.60 - Quantum gravity.\\ 

\bigskip
\bigskip
{\bf 1. Introduction}\\
\vskip 2mm

As recently emphasized by Poppitz and Trivedi \cite{pt}, from the point of
view of the concept of supersymmetry breaking the difference between
supersymmetric quantum mechanics (SUSYQM) and the 3+1 dimensional
renormalizable
supersymmetric field theory with spin-0 and spin-1/2 fields is only minimal
in the sense that the ``spin-orbit" term in the first case corresponds to the
Yukawa interaction
between the bosons and fermions in the supermultiplet in the second case.
This can also be seen
by noting that the N=1 four-dimensional SUSY algebra \cite{book}
$$
\{ Q_{\alpha},\bar Q_{\dot \alpha}\}=-2\sigma_{\alpha, \dot \alpha}^{\mu}
P_{\mu}
\eqno(1)
$$
reduces, in the rest frame of the system, $P_0=H$, $\vec{P}=0$, after
appropriate scaling, to the nonrelativistic SUSYQM algebra. Thus, it might
be possible that lessons from the simple (toy) SUSYQM be useful when
extrapolated in an unambiguous way to
the complicated field theories. As a matter of fact, at the
conceptual level, one may well make use of toy models for better pointing
out the main ideas. Here, the toy (simple) idea is a double Darboux,
strictly isospectral
scheme of unbroken SUSYQM first discussed in the 80's by
Mielnik \cite{M} but
not so well known even to people working in SUSYQM (for recent applications
see \cite{roa}).
In this work, the acronym DDGR (double Darboux general Riccati) will be used
for this mathematical scheme.
The key feature of the DDGR method is to introduce a sort of
free parameters labeling the DDGR-modulated degenerate vacua.

Usually, in quantum field theories the full unbroken supersymmetry is
considered as an uninteresting case. The favored idea is that of
supersymmetry unbroken
at the tree level, but broken due to instanton tunneling, because the
instanton
calculus can generate small scales therefore explaining the hierarchy of
scales in nature. Moreover, the Euclidean instanton calculus is also basic
in the fundamental topic of the quantum creation of the universe.
However, within the full unbroken nonrelativistic SUSY, the DDGR construction
can produce degenerate true zero modes, that may be considered as
nonrelativistic counterparts of quantum field degenerate vacua. This important
fact is practically unknown to field theorists.

The organization of the paper is the following.
In the next section, I provide a short but detailed sketch of the DDGR
scheme and present its main ingredients.
In section 3, I comment on the application to minisuperspace
quantum gravity/cosmology where the appropriate
DDGR parameter can be determined from a stationary phase approximation
and I end up with conclusion.

\newpage
\bigskip
\bigskip
{\bf 2. The DDGR construction}\\
\vskip 2mm

Witten's SUSYQM \cite{W}
looks so simple that apparently is well known to every layman and will not be
repeated here.
There is however more than that as first shown by Mielnik \cite{M}
for the harmonic
oscillator case. It is possible to construct families of potentials
{\em strictly} isospectral with respect to the initial (bosonic) one, if
one asks for the most general superpotential (i.e., the general
Riccati solution)
such that $\rm V_+(x)=  w_{g}^2 + \frac{d w_{g}}{dx}$, where ${\rm V_+}$ is
the fermionic partner potential. It is easy to see that
one particular solution to this equation is ${\rm w_p= w(x)}$, where w(x) is
the common Witten superpotential. One is led to consider the following
Riccati equation ${\rm  w_{g}^2 + \frac{d w_{g}}{dx}=w^2_p +\frac{d w_p}{dx}}$,
whose general solution can be written down as 
${\rm w_{g}(x)= w_p(x) + \frac{1}{v(x)}}$, where ${\rm v(x)}$ is an unknown
function. Using this ansatz, one obtains for the function ${\rm v(x)}$ the
following Bernoulli equation
$$
{\rm \frac{dv(x)}{dx} - 2 \, v(x)\, w_p(x) = 1},
\eqno(2)
$$
that has the solution
$$
{\rm v(x)= \frac{{\cal I}_0(x)+ \lambda}{u_{0}^{2}(x)}},
\eqno(3)
$$
where ${\rm {\cal I}_0(x)= \int^x \, u_0^2(y)\, dy}$, and $\lambda$ is an
integration constant thereby considered as a free DDGR parameter.
Thus, ${\rm w_{g}(x)}$ can be written as follows
$$
{\rm w_{g}(x;\lambda)=  w_p(x) + \rm \frac{d}{dx}} \Big[ {\rm ln}
({\cal I}_0(x) + \lambda) \Big]
\eqno(4a)
$$
$$
={\rm w_p(x)+\sigma _{0}(\lambda)}
\eqno(4b)
$$
$$
={\rm - \frac{d}{dx} \Big[ ln \left(\frac{u_0(x)}{{\cal I}_0(x) +
\lambda}\right)\Big]}.
\eqno(4c)
$$
Finally, one easily gets the $V_-(x;\lambda)$ family of
potentials 
$$
{\rm  V_-(x;\lambda)} = {\rm w_{g}^2(x;\lambda) -
\frac{d w_{g}(x;\lambda)}{dx}}
\eqno(5a)
$$
$$
= {\rm V_-(x) - 2 \frac{d^2}{dx^2} \Big[ ln({\cal I}_0(x) + \lambda)}
\Big]
\eqno(5b)
$$
$$
= {\rm V_-(x) -2\sigma _{0,x}(\lambda)}
\eqno(5c)
$$
$$
= {\rm V_-(x) - \frac{4 u_0(x) u_0^\prime (x)}{{\cal I}_0(x)
+ \lambda} 
+ \frac{2 u_0^4(x)}{({\cal I}_0(x) + \lambda)^2}.}
\eqno(5d)
$$
All ${\rm  V_-(x;\lambda)}$ have the same supersymmetric partner potential
${\rm V_+(x)}$ obtained by deleting the ground state.
They are asymmetric double-well potentials that may be considered as a sort
of intermediates between the bosonic potential ${\rm V_-(x)}$ and
the fermionic partner ${\rm V_+(x)=V_-(x)-2\sigma _{0,x}(x)}$, where
${\rm \sigma _{0,x}(x)=\frac{d^2}{dx^2}ln u_0}$, is the
notation for logarithmic derivatives in the book of
Matveev and Salle \cite{MS}.
From Eq. ($4c$) one can infer the ground state wave functions
for the potentials ${\rm V_-(x;\lambda)}$ as follows
$$
{\rm u_0(x;\lambda)= f(\lambda)
\frac{u _0(x)}{{\cal I}_0(x) + \lambda}},
\eqno(6)
$$
where ${\rm f(\lambda)}$ is a normalization factor that can be shown to be
of the form
${\rm f(\lambda)= \sqrt{\lambda(\lambda +1)}}$.
One can now understand the double Darboux feature of the strictly isospectral
construction by writing the parametric family in terms of their unique
``fermionic" partner
$$
{\rm V_{-} (x;\lambda)=V_{+}(x)
-2\frac{d^2}{dx^2}\ln\left(\frac{1}{u_{0}(x;\lambda)}\right)},
\eqno(7)
$$
which shows that the DDGR transformation is of the
inverse Darboux type \cite{MS1}, allowing at the same time
a two-step (double Darboux) interpretation,
namely, in the first step one goes to the fermionic system and in the
second step one returns to a deformed bosonic system.

From the normalization factor one can see that in the $\lambda$-parameter
space the interval [-1,0] is forbidden. A connection with
other isospectral methods has been found, by noticing that
the limiting values -1 and 0 for the parameter $\lambda$ lead to the
Abraham-Moses procedure
\cite {AbMo} and Pursey's one \cite {Pu}, respectively. Actually, the
discussion
is more involved because of the singularities that may appear both in
the strictly isospectral solutions and potentials, see \cite{brssv}.
If the normalization
factor $f(\lambda)$ is included all ${\rm u_0(x;\lambda)}$ are true
Schroedinger
zero modes labeled by $\lambda$. According to our experience \cite{roa}, the
denominator of these
DDGR zero modes acts as a modulational factor, introducing
some additional structure in the shape of the zero mode as a result of the
form of the DDGR double-well potentials.
More details on the construction such as the
connection with the general zero-energy Schroedinger solution and an
intertwining operator approach can be
found in my recent work in collaboration \cite{brssv}.
The DDGR method can be applied to any one-dimensional
system, whose dynamics is dictated by a Schroedinger(-like) equation.
Moreover, one can employ combinations of any pairs of
Abraham-Moses procedure, Pursey's one, and the Darboux one.
However, only the DDGR method leads to reflection and
transmission amplitudes identical to those of the original potential, showing
the complete degeneracy produced by such a construction. Moreover, the scheme
can be used iteratively leading to multiple-parameter families of
solutions \cite{psp,brssv}.

Now, let us present some small steps toward extrapolating DDGR to much more
complicated theories. In my view, if this will prove possible,
the following philosophy is suggested by the DDGR method.

(i)
The fermionic system, as it stands in the
Schroedinger-Riccati ``entanglement" \cite{entan},
is merely an intermediate step of the mathematical procedure.

(ii)
If one assigns physical meaning to the DDGR vacua,
since they are parametrically defined, one may think of fixing the parameter
(and thus the bosonic vacuum) through some mathematical procedure.
In the next section, for the example of minisuperspace cosmology, it is
briefly shown how one can fix the parameter through a stationary phase
approximation.

(iii)
One can also claim that an extrapolation of the
DDGR construction to gauge theories might offer a different perspective on
high energy physics, allowing the freedom of
selecting the true bosonic vacuum from a DDGR parametric family of vacua.

\bigskip
\bigskip
{\bf 3. DDGR methods in quantum cosmology}\\
\vskip 2mm

It is well known that any system invariant under spacetime
reparametrizations has a vanishing Hamiltonian; for a compact discussion
the reader is directed to a paper by Gamboa and Zanelli \cite{gz}.
Such a situation is common in quantum gravity and cosmology \cite{Mar}.
Quantum cosmology is an area dominated by the Wheeler-DeWitt (WDW) equation.
In some simple, minisuperspace
models the WDW equation can be reduced to a stationary
Schroedinger-like equation at zero energy.
In papers with Socorro, we have already applied DDGR to a couple of
minisuperspace models \cite{roso}.
Generically, a WDW equation of DDGR type can be
written down as follows
$$
{\rm -\frac{d^2\Psi (\Omega;\lambda)}{d\Omega ^2}+V(\Omega,\lambda)
\Psi (\Omega; \lambda)}=0,
\eqno(8)
$$
where, as an example,
$\Omega$ is Misner's minisuperspace variable \cite{Mis}, and
$\Psi (\Omega;\lambda)$ are solutions of the type given by Eq. (6), either
with
or without the normalization factor. Here I pose the problem of what would
be an appropriate $\lambda$ in quantum cosmology.
To answer this question, I recall that Salopek \cite{sal}
discussed an interesting (semiclassical) principle of superposition for
Hamilton-Jacobi (HJ) theory that applies to Schroedinger solutions which
depend on a continuous parameter.
For Schroedinger solutions
${\rm \psi (x;\lambda)}$
depending on a continuous parameter $\lambda$ any linear superposition
is also a solution
$$
{\rm \psi(x)=\int d\lambda p(\lambda)\psi(x;\lambda)},
\eqno(9)
$$
where the weighting function ${\rm p(\lambda)}$ is arbitrary.
If we work in the
semiclassical limit, $\hbar\rightarrow 0$, then ${\rm \psi(x;\lambda)}$ and
${\rm p(\lambda)}$ may be approximated by phase factors
$$
{\rm \psi(x;\lambda)\approx e^{iS(x;\lambda)/\hbar}~, \quad \quad
p(\lambda)=
e^{ig(\lambda)/\hbar}}~.
\eqno(10)
$$
${\rm S(x;\lambda)}$ is then a solution of the HJ equation which depends on the
parameter $\lambda$. As remarked by Salopek, if ${\rm S}$ is real, then one
deals with classical phenomena, whereas if ${\rm S}$ is complex, one may
describe quantum phenomena such as tunneling or the initial wavefunction
of the universe.
The superposition integral may be approximated using the
stationary phase approximation
$$
{\rm \psi(x)=\exp [i(S(x;\lambda _{st})+g(\lambda _{st}))/\hbar]},
\eqno(11)
$$
where ${\rm \lambda _{st}=\lambda (x)}$ is now chosen so that the phase of
the integrand has a maximum or minimum, i.e.,
$$
{\rm \frac{\partial}{\partial \lambda}\Big[S(x;\lambda)+g(\lambda)\Big]=0},
\eqno(12)
$$
for ${\rm \lambda =\lambda _{st}}$.

To implement these ideas within the DDGR solutions
one should consider the superposition
$$
{\rm \Phi _{0}(\Omega)=\int d\lambda p(\lambda)\Psi _{0}(\Omega;\lambda)},
\eqno(13)
$$
as the most appropriate WDW cosmological solution
and apply a stationary phase approximation as above.
Since the DDGR solutions are of
the type ${\rm \Psi _{0}(\Omega;\lambda)\propto
e^{-\int ^{\Omega}w_{g}dy}}$,
one gets ${\rm S\propto -i\hbar\int ^{\Omega}w_{g}dy}$,
i.e, ${\rm S\propto -i\hbar \Psi _{0}(\Omega;\lambda)}$ and the condition of
stationary phase approximation reads
$$
i{\rm \frac{\partial}{\partial \lambda}\Big[-\Psi _{0}(\Omega;\lambda)+
g(\lambda)\Big]=0},
\eqno(14)
$$
from which one can get ${\rm \lambda _{st}(\Omega)}$.

\bigskip
\bigskip
{\bf 4. Conclusion}\\
\vskip 2mm

I have tackled the toy idea of DDGR
supersymmetric scheme, presenting its main ingredients. It might be that,
as suggested in this work, an unambiguous extrapolation of this construction
to relativistic (gauge)
physics, if possible, may show up interesting consequences, such as parametric
families of DDGR bosonic field vacua. As commented, for the case of
minisuperspace quantum cosmology, one can determine the appropriate DDGR
cosmological parameter from a stationary phase approximation.

\bigskip
\bigskip
\noindent
This work was supported in part by the CONACyT project 458100-5-25844E.
The author wishes to thank the referee for his comments.


\end{document}